\documentstyle[preprint,prl,aps]{revtex}

\tighten

\def\r{{\bf r}}
\def\px{{\partial \over \partial x}}
\def\pt{{\partial \over \partial t}}
\def\bra{\langle}  \def\ket{\rangle}

\def\hp{\hat \psi}
\def\efn{\hat \psi (\r,t)} \def\efd{\hat \psi^\dagger (\r,t)}
\def\cfn{\hat \psi_C (\r,t)} 
\def\rfn{\hat \psi_R (\r,t)} 
\def\ofn{\hat \psi_1 (x,t)} \def\ofd{\hat \psi_1^\dagger (x,t)}

\begin{document}
\title{
Interacting One-Dimensional Electrons \\
Driven by Two-Dimensional Reservoir Electrons
}
\author{Akira Shimizu$^*$ and Takayuki Miyadera$^{**}$}
\address{Institute of Physics, University of Tokyo, 
3-8-1 Komaba, Meguro-ku, Tokyo 153, Japan \\
Fax: +81-3-5790-7229, E-mail: shmz@ASone.c.u-tokyo.ac.jp,$^*$  
miyadera@ASone.c.u-tokyo.ac.jp$^{**}$
}
\date{21 August 1997}
\maketitle
\medskip

\begin{center}
{\bf Abstract}
\end{center}

We derive an effective 1D theory 
from the Hamiltonian of the 3D system which consists of 
a mesoscopic conductor and reservoirs.
We assume that the many-body interaction have the same magnitude in 
the conductor as that in the reservoirs, in contrast to 
the previous theories which made the ad hoc assumption that
the many-body interaction were absent in the reservoirs. 
We show the following:
(i) The effective potentials of impurities and two-body 
interaction for the 1D modes become weaker as $x$ goes away from 
the conductor. 
(ii) On the other hand, 
the interaction between the 1D and the reservoir modes is important in the reservoir regions, where 
the reservoir modes excite and attenuate the 1D modes 
through the interaction.
(iii) As a result, 
the current $\hat I_1$ of the 1D modes is not conserved, 
whereas the total current $\hat I$ is of course 
conserved.
(iv) For any steady state 
the total current $\bra I \ket$, its equilibrium fluctuation 
$\bra \delta I^2 \ket^{eq}$
at low frequency, and non-equilibrium fluctuation 
$\bra \delta I^2 \ket^{noneq}$ at low frequency, 
of the original system are independent of $x$, 
whereas $\bra \delta I^2 \ket^{eq}$
and $\bra \delta I^2 \ket^{noneq}$ at higher  
frequencies may depend on $x$. 
(v) Utilizing this property, 
we can evaluate $\bra I \ket$, $\bra \delta I^2 \ket^{eq}$, 
and $\bra \delta I^2 \ket^{noneq}$ at low frequency from 
those of the 1D current $\hat I_1$.
(vi) In general, the transmittance $T$ in the Landauer formula should be 
evaluated from a single-body Hamiltonian which includes  
a Hartree potential created by the density deformation which is 
caused by the external bias.


\newpage
\bigskip\noindent{\bf 
1.\ Identification of the single-body part of the 3D Hamiltonian
}\medskip

We consider a system which consists of a mesoscopic conductor 
and reservoirs (Fig.\ 1), in which  
electrons are interacting each other according to the 
three-dimensional (3D) Hamiltonian (in the Heisenberg picture),
\begin{eqnarray}
&&
\hat H 
= 
\hat H^0_0 + \hat V_0
\label{H01}\\
&& \quad
=
\int d^3r \ \hat {\cal H}^0_0(\r,t)
+
{1 \over 2}
\int d^3r \int d^3r' \ \hat {\cal V}_0(\r,\r',t),
\label{H02}\\
&&
\hat {\cal H}^0_0(\r,t) 
=
\efd \left[
- {\hbar^2 \over 2 m} \nabla^2 + u^c(\r)+u^i(\r) + e \phi_{ext}(\r)
\right] \efn,
\\
&&
\hat {\cal V}_0(\r,\r',t)
=
\hat \rho(\r,t)v(\r - \r') \hat \rho(\r',t),
\label{H04}\end{eqnarray}
where $\efn$ is the 3D electron field, 
$u^c$ the confining potential, $u^i$ the potential of impurities
and/or defects,
$\phi_{ext}$ the external electrostatic potential,  
and $v(\r - \r') \equiv e^2/|\r-\r'|$ denotes the Coulomb potential  
between the local charge density 
$
\hat \rho(\r,t) 
\equiv
\efd \efn - \rho_{BG}(\r)
$.
Here, $e \efd \efn$ is the charge density of the electrons and 
$-e \rho_{BG}$ is that of the background charges which are 
assumed to be fixed.
Note that we cannot simply add  to $\hat H$ 
the position-dependent ``chemical potential 
term" like $-\int d^3r \ \mu(\r) \efd \efn$, 
because it does not commute with $\hat H$. 

We are interested in a steady state, which may be either equilibrium or 
non-equilibrium. Its density operator is denoted by $\hat \zeta$, 
which is to be determined self-consistently. 
The average charge density of this state is 
$
{\rm Tr} \left[ \hat \zeta e \hat \rho(\r,t)\right] 
\equiv e \rho_{av}(\r)
$,
which is a function of the bias voltage.
Although $\rho_{av}(\r)$ vanishes at $r \to \infty$,  
it is finite and $\r$ dependent in finite regions of space, 
because 
mesoscopic systems (conductor plus reservoirs) are spatially inhomogeneous.
The finite $\rho_{av}$ induces a long-range effect, 
which does not vanish by the screening, through 
the Coulomb interaction. 
To construct a meaningful 1D theory,
in which the interaction of a ``1D field" (Eq.\ (\ref{1Df})) 
with a ``reservoir field" (Eq.\ (\ref{Rf})) is not strong, 
we must remove the long-range effect from the interaction Hamiltonian.
This can be accomplished 
by 
rewriting Eqs.\ (\ref{H01})-(\ref{H04}) as
\begin{eqnarray}
&&
\hat H
=
\hat H^0 + \hat V + V_{av}
\label{H1}\\
&& \quad
=
\int d^3r \ \hat {\cal H}^0(\r,t)
+
{1 \over 2}
\int d^3r \int d^3r' \left[ \hat {\cal V}(\r,\r',t)
+
{\cal V}_{av}(\r,\r')
\right],
\label{H2}\\
&&
\hat {\cal H}^0(\r,t) 
=
\efd \left[- {\hbar^2 \over 2 m} \nabla^2 
+ u^c(\r)+u^i(\r)+ e\phi_{av}(\r) \right] \efn,
\label{Hfree}\\
&&
\hat {\cal V}(\r,\r',t)
=
\delta \hat \rho(\r,t)v(\r - \r') \delta \hat \rho(\r',t),
\\
&&
{\cal V}_{av}(\r,\r')
=
- \rho_{av}(\r)v(\r - \r') \rho_{av}(\r').
\label{H5}\end{eqnarray}
Here, 
$
\delta \hat \rho(\r,t)
\equiv
\hat \rho(\r,t) - \rho_{av}(\r)
$
denotes the charge fluctuation around $\rho_{av}$, and 
$\phi_{av}(\r)$ is the average electrostatic potential;
\begin{equation}
e \phi_{av}(\r) \equiv \int d^3r' v(\r-\r') \rho_{av}(\r') 
+ e \phi_{ext}(\r).
\label{pav}\end{equation}
The advantage of rewriting $\hat H$ in the form of 
Eqs.\ (\ref{H1})-(\ref{pav}) is that 
effects of $\hat V$ is much weaker than those of the original 
interaction $\hat V_0$. In particular, $\hat V$ does not 
cause the long-range effect because it is already included in $\hat H^0$.
It is therefore much better to start with these equations, taking 
$\hat H^0$ (rather than $\hat H_0^0$) as the single-body part.
For example, 
we can derive the Landauer formula by neglecting 
$\hat V$ in Eq.\ (\ref{H1}); 
\begin{equation}
G = (2 e^2 /h) T[\phi_{av}].
\label{LF}\end{equation}
Here, the transmittance $T[\phi_{av}]$ is calculated from 
$\hat {\cal H}^0$, Eq.\ (\ref{Hfree}), which 
includes the self-consistent potential $\phi_{av}$. 
On the other hand, if we neglected 
$\hat V_0$ in Eq.\ (\ref{H01}), we would obtain 
$G = ({2 e^2 / h}) T[\phi_{av}=0]$.
This is different from Eq.\ (\ref{LF}) when the bias voltage is finite.
Since actual measurements on mesoscopic conductors and 
on the quantum Hall effect are often 
performed under a non-negligible bias, 
one should use Eq.\ (\ref{LF}) for these experiments.

\bigskip\noindent{\bf 
2.\ Decomposition of the electron field 
}\medskip

Using the above form of the Hamiltonian, we now 
decompose the electron field into two parts.
We are interested in the case where 
the variation of $W(x)$ of Fig.1 is slow,  
{\it i.e.}, $W(x) \simeq W(x+\lambda_F)$, where $\lambda_F$ is 
a length of the order of the 
Fermi wavelength.
Otherwise,
undesirable reflections would occur at the boundary regions between 
the conductor and reservoirs, and experimental results for such samples 
would not be very interesting.
We do not want reflections either in the conductor except 
for reflections by impurities and/or defects.
If the bias voltage is not too high, these conditions 
can be stated as the assumption that 
the $x$ dependence of $u^c(\r)+e \phi_{av}(\r) \equiv u(\r)$ is weak,
{\it i.e.}, $u(x,y,z) \simeq u(x+\lambda_F,y,z)$.
We also assume that 
only the lowest subband (see Eq.\ (\ref{sb}) below) 
is occupied by electrons.

Under these assumptions 
we can find approximate solutions of the single-body Schr\"odinger equation 
\begin{equation}
\left[- {\hbar^2 \over 2 m} \nabla^2 + u(\r) \right] \varphi(\r)
=
\varepsilon \varphi(\r),
\label{se}\end{equation}
for $\varepsilon$ not far from the Fermi energy, in the following form:
\begin{equation}
\varphi_{k}(\r) \simeq 
{1 \over \sqrt{\cal L}}
\exp \left[ i \int_0^x K_{k}(x) dx \right]
\varphi^\bot(y,z;x),
\end{equation}
where ${\cal L}$ is the normalization length in the $x$ direction, and 
$\varphi^\bot(y,z;x)$ 
is the normalized 
eigenfunction belonging to the lowest eigenvalue 
$u^\bot(x)$ of the following eigenvalue 
equation (in which $x$ is regarded as a parameter):
\begin{equation}
\left[- {\hbar^2 \over 2 m} 
\left( 
{\partial^2 \over \partial y^2} + {\partial^2 \over \partial z^2}
\right)
+ u(\r) \right] \varphi^\bot(y,z;x)
=
u^\bot(x) \varphi^\bot(y,z;x).
\label{sb}\end{equation}
The $x$-dependent wavenumber $K_{k}(x)$ is defined by
$
\varepsilon_{k} = 
\hbar^2 K_{k}(x)^2/2m + u^\bot(x)
$,
where $\varepsilon_{k}$ denotes the eigenenergy of the state 
$\varphi_{k}(\r)$. 
The $\varepsilon_{k}$ can be expressed as 
$
\varepsilon_{k} = 
\hbar^2 k^2/2m + \varepsilon^\bot
$, where
$k \equiv K_{k}(0)$ and $\varepsilon^\bot \equiv u^\bot(0)$
are the wavenumber and the lowest subband energy, respectively,  
at the center of the conductor.
We call $\varphi_k$'s ``conductor modes".
There are other solutions to Eq.\ (\ref{se}), 
which are denoted by $\varphi_\nu $.
They include 
states whose $\varepsilon$ is far from the Fermi energy, 
states belonging to higher subbands, and 
states which decay exponentially in the conductor.
We call these solutions ``reservoir modes." 
The $\r$ dependence of the electron field $\efn$ can be 
expanded in terms of 
$\varphi_k$'s and $\varphi_\nu$'s.
We can thus decompose the electron field into two parts as
$ 
\efn = \cfn + \rfn,
$ 
where 
\begin{eqnarray}
\cfn 
&\equiv& 
\sum_k \varphi_k(\r) \int d^3r' \varphi_k^*(\r') \hp (\r',t)
\equiv
\sum_k \varphi_k(\r) \hat c_k(t),
\label{Cf}\\
\rfn 
&\equiv& 
\sum_\nu \varphi_\nu(\r) \int d^3r' \varphi_\nu^*(\r') \hp (\r',t)
\equiv
\sum_\nu \varphi_\nu(\r) \hat d_\nu(t).
\label{Rf}\end{eqnarray}
We call $\hp_R$ the ``reservoir field."

We now construct a 1D field $\ofn$ from the 3D field $\cfn$ as
\begin{equation}
\ofn 
\equiv 
\int \int dy dz \ \varphi^{\bot *}(y,z;x) \ \cfn  
\simeq 
\sum_k
{1 \over \sqrt{\cal L}}
\exp \left[ i \int_0^x K_{k}(x) dx \right] \hat c_k(t).
\label{1Df}\end{equation}
This 1D field has the same contents as $\hp_C$, because the inverse transformation can be accomplished as
$\cfn = \varphi^\bot(y,z;x) \ofn$.
In terms of $\hp_1$ and $\hp_R$, the Hamiltonian can be recast as
\begin{equation}
\hat H 
= 
\hat H_C^0 + \hat V_{C} 
+ \hat V_{CR} + \hat H_R^0 + \hat V_{R} +V_{av}
\label{HCR}\end{equation}
Here, 
$\hat H_C^0$ ($\hat H_R^0$) denotes the single-body part 
of the 1D (reservoir) field, $\hat V_{C}$ ($\hat V_{R}$) is its mutual interaction, and 
$\hat V_{CR}$ denotes the interaction between the 1D and reservoir fields.
In particular, 
$\hat H_C^0 = \int dx \ \hat {\cal H}_C^0(x,t)$ 
and
$\hat V_{C} = {1 \over 2}
\int dx \int dx' 
\hat {\cal V}_{C}(x,x',t)
$,
where,  
\begin{eqnarray}
\hat {\cal H}_C^0(x,t)
&\simeq&
\ofd \left[ 
- {\hbar^2 \over 2m} {\partial^2 \over \partial x^2} 
+ u^\bot(x) +u^i_1(x) 
\right] \ofn,
\\
\hat {\cal V}_{C}(x,x',t)
&=&
\delta \hat \rho_1(x,t)v_1(x,x') \delta \hat \rho_1(x',t).
\end{eqnarray}
Here,  
$\delta \hat \rho_1(x,t) \equiv \ofd \ofn - \bra \ofd \ofn \ket$
is the density fluctuation of the 1D field, and
\begin{eqnarray}
u^i_1(x)
&\equiv&
\int \int dy dz \ |\varphi^\bot(y,z;x)|^2 \ u^i(\r),
\\
v_1(x,x')
&\equiv&
\int \int dy dz \int \int dy' dz' \ 
|\varphi^\bot(y,z;x)|^2 \ v(\r-\r') \ 
|\varphi^\bot(y',z';x')|^2.
\label{HCR5}\end{eqnarray}

The 3D impurity potential $u^i(\r)$ is 
spatially averaged, with the weight $|\varphi^\bot|^2$, 
to give the effective 1D impurity potential $u^i_1(x)$.
In the conductor region, $u^i_1(x)$ has the same order of magnitude 
as $u^i(\r)$.
In the reservoir regions, on the other hand, 
$u^i_1(x)$ becomes smaller because 
$\varphi^\bot$ extends over the width $W(x)$ in the $y$ direction, 
and $u^i_1(x)$ is the averaged value of the random variable $u^i(\r)$ 
over this wide area.
As $x \to \pm \infty$, in particular, $u^i_1(x) \to 0$ if $W(x) \to \infty$.

We can say the same for the two-body interaction potential 
$v_1(x,x')$.
In the conductor region, the major effect of 
the integrals over $y,z,y',z'$ of Eq.\ ({\ref{HCR5}) 
is simply to smear out the singularity 
of $v(\r - \r')$ at $\r=\r'$ 
by averaging over the conductor width.
In the reservoir regions, 
on the other hand, 
the interaction becomes extremely weak as we show now.
When $x$ or $x'$ is in a reservoir region(s), 
$v_1(x,x')$ 
is screened by the reservoir modes through $\hat V_{CR}$.
It is therefore better to rewrite 
$\hat V_{C} + \hat V_{CR}$ in Eq.\ (\ref{HCR}) as 
$\hat V_{C}^{sc} + \hat V_{CR}'$, 
where 
$\hat V_{C}^{sc}$ is the screened, short-range interaction and 
$\hat V_{CR}' \equiv \hat V_{CR} + \hat V_{C} - \hat V_{C}^{sc}$.
For large $|x - x'|$ it is clear that 
$\hat V_{C}^{sc}$ becomes vanishingly weak as $|x - x'|$ is increased.
For $x \sim x'$, on the other hand, 
the weakness of $\hat V_{C}^{sc}$ is 
never trivial because 
the screening is ineffective at a short distance.
To investigate this case, 
we consider the behavior of the original, 
long-range potential $v_1(x,x')$.
It is easy to show that 
$v_1(x,x') \propto 1/W(x)$ as $x \sim x' \to \pm \infty$. 
Hence $v_1 \to 0$ if $W(x) \to \infty$.
Since the interaction potential of $\hat V_{C}^{sc}$ 
decays more quickly than $v_1$, 
it also approaches zero, more quickly, 
as $x \sim x' \to \pm \infty$.

We have thus shown that the potentials of both the impurities  
and two-body interaction for the 1D field vanish
as $x$ and/or $x'$ go away from the conductor.
(It should be stressed that 
$W(x) \to \infty$ is necessary 
for this property; otherwise, the impurity potential and 
the short range part of the interaction 
would remain finite in all regions.)
This may partly justify the 1D models [2] 
in which a 1D field is assumed to become free as $x,x' \to \pm \infty$.
However, note that $\hat V_{CR}'$ will cause 
additional effects on the 1D field: 
the reserver modes excite, attenuate, and renormalize 
the 1D field [1].

\bigskip\noindent{\bf 
3.\ Calculation of the total current from the 1D current
}\medskip

We now turn to observables.
What is measured in most experiments is 
the current $\hat I$ which is given by 
$ 
\hat I(x,t) \equiv \int \int dy dz 
\hat J_x (\r,t),
$ 
where $\hat J_x$ denotes the $x$ component of the current density.
From the continuity equation for $\hat {\bf J}$ and $\hat \rho$, 
we have
\begin{equation}
\px \hat I (x,t) + \pt \hat \Lambda (x,t) =0,
\label{ceI}
\end{equation}
where $\hat \Lambda$ is the 1D density of electrical charge;
$
\hat \Lambda(x,t) 
\equiv 
\int \int dy dz \ e \hat \rho (\r,t)
$.
For a steady state, either equilibrium or non-equilibrium, 
Eq.\ (\ref{ceI}) yields 
$
\px \bra \hat I (x,t) \ket 
= -\pt \bra \hat \Lambda (x,t) \ket=0
$.
That is, the average current $\bra \hat I (x,t) \ket$ 
is independent of $x$.

Note, however, that this is not the case for the current fluctuations.
For example, consider the symmetrized correlation function of
$\delta \hat I(x,t) \equiv \hat I(x,t) - \bra \hat I(x,t) \ket$,
\begin{equation}
C_{II}(x,t-t') \equiv 
\bra \delta \hat I(x,t) \delta \hat I(x,t') 
+ \delta \hat I(x,t') \delta \hat I(x,t)
\ket/2,
\end{equation}
which may be taken as a definition of the current fluctuation. 
Equation (\ref{ceI}) and the time-translational invariance yield
\begin{equation}
\px C_{II}(x,t)
=
\pt
\left[
C_{I \Lambda}(x,t) + C_{\Lambda I}(x,-t)
-C_{I \Lambda}(x,-t) - C_{\Lambda I}(x,t)
\right],
\label{ceC}
\end{equation}
where we have introduced the (non-symmetrized) density-current correlation,
$
C_{I\Lambda}(x,t-t') \equiv 
\bra \delta \hat I(x,t) \hat \Lambda (x,t') \ket
$ and 
$
C_{\Lambda I}(x,t-t') \equiv 
\bra \hat \Lambda (x,t) \delta \hat I(x,t') \ket
$.
It is seen that {\it the current fluctuation
generally depends on the position $x$.}

However, we can show that the low-frequency component of 
its spectral intensity,
$
\bra \delta I^2 \ket_\omega
\equiv
\int_{-\infty}^{\infty}
e^{i \omega t} C_{II}(x,t)
dt
$,
is independent of $x$.
To show this, we note that 
any real systems which possess the thermodynamic stability must 
have the ``mixing property"; namely, any correlation functions 
decay as $|t-t'| \to \infty$.
This property might be lost if one took  
an oversimplified model, such as integrable models, for $\hat H$.
However, our 
$\hat H$ contains both the conductor and reservoir modes
as well as their interactions.
It is quite reasonable to assume that 
such a complicated 3D model 
has the mixing property.
Therefore, $C_{II}(t)$ vanishes as t exceeds a 
``correlation time" $\tau_c$. Then, for $\omega \ll 1/\tau_C$ 
we find from Eq.\ (\ref{ceC}) 
\begin{equation}
\px \bra \delta I^2 \ket_{\omega}
\simeq
\px \bra \delta I^2 \ket_{\omega=0}
=
2 \left[
C_{I \Lambda}(x,\infty) + C_{\Lambda I}(x,-\infty)
-C_{I \Lambda}(x,-\infty) - C_{\Lambda I}(x,\infty)
\right].
\label{xd}\end{equation}
The right-hand side vanishes because of the mixing property.
We have thus obtained the theorem:
{\it For a steady state, 
the average current 
is independent of $x$, and 
the low-frequency ($\omega \ll 1/\tau_c$) component of 
the spectral intensity of current fluctuation is also independent of $x$, 
whereas higher-frequency components may depend on $x$.}
This theorem is quite general: 
It applies to any steady states, 
either equilibrium or nonequilibrium, of real systems and of 
any theoretical models which are realistic enough 
so that they have the mixing property.

The theorem is very useful for our effective 1D theory.
Let us define the current of the 1D field by
\begin{equation}
\hat I_1(x,t) 
\equiv
{e \over 2 m} \left[
\ofd \left\{ {\hbar \over i}{\partial \over \partial x} \ofn 
\right\} + {\rm h.c.}
\right].
\end{equation}
Note that this 1D current is {\it not} conserved because of the interaction 
$\hat V_{CR}$, which transforms the 1D current into  
the reservoir current and vice versa.
At first sight, this might seem to cause difficulties.
Fortunately, 
however, 
the above theorem guarantees that we can evaluate 
$\bra I \ket$ and $\bra \delta I^2 \ket_{\omega \simeq 0}$ at arbitrary $x$.
If we take $x=0$ (the center of the conductor), 
$\hat I(x,t) \simeq \hat I_1(x,t)$ because 
the reservoir field does not contribute there.
Therefore, we can evaluate 
$\bra I \ket$ and $\bra \delta I^2 \ket_{\omega \simeq 0}$ by 
evaluating 
$\bra I_1 \ket$ and $\bra \delta I_1^2 \ket_{\omega \simeq 0}$ at $x=0$.
This may partly justify the 
Landauer-like approach [3], 
in which one evaluates the DC conductance and nonequilibrium 
current fluctuation by evaluating 
$\bra I_1 \ket$ and $\bra \delta I_1^2 \ket_{\omega \simeq 0}$ 
using an effective 1D model.

We have thus shown that it is sufficient to evaluate the 1D current 
$\hat I_1$.
Therefore, we can project out the reservoir degrees of freedom, 
which should behave as the 
usual two- or three-dimensional electrons because 
the reservoirs are large, 
in the standard manner [4].
By doing so, we can derive either the generalized Langevin equation [4]
for $\hat I_1$, or the Liouville equation 
for the reduced density operator [4], 
$\hat \zeta_C \equiv {\rm Tr}_R [\hat \zeta]$,
where the trace is taken over the Hilbert space of the reservor field.
The original 3D model of the conductor plus reservoirs has
thus been reduced to the 1D model.
Detailed discussions and applications to various problems will be 
described elsewhere [1].

This work has been supported by the Core Research for Evolutional Science
and Technology (CREST) of the Japan Science and Technology Corporation (JST).
\medskip
\parindent 0mm

\begin{center}
{\bf References}
\end{center}

[1] A. Shimizu and T. Miyadera, unpublished.

[2] For example, A. Shimizu and M. Ueda,
Phys. Rev. Lett. {\bf 69} (1992) 1403;
D.L. Maslov and M. Stone, 
Phys. Rev. B {\bf 52} (1995) R5539;
V.V. Ponomarenko, {\it ibid}, R8666;
I. Safi and H.J. Schulz, {\it ibid}, R17040.

[3] A. Shimizu, J. Phys. Soc. Jpn. {\bf 65} (1996) 1162 
(Errata, {\it ibid}, 3096).

[4] R. Kubo, M. Toda and N. Hashitsume, {\it Statistical Physics II}, 2nd ed. (Springer, 1992).

%\newpage

\begin{center}
{\bf Figure caption}
\end{center}
\medskip

Fig.1 \
The mesoscopic conductor and reservoirs.

\end{document}